\documentclass{JFM-FLM-arxiv}
\usepackage{graphicx}
\usepackage{newtxtext}
\usepackage{newtxmath}
\usepackage{natbib}
\usepackage{hyperref}

\usepackage{bm}

\title{\fontsize{14pt}{17pt} \selectfont Forcing regimes in the two-dimensional Navier-Stokes equations}

\author{Ritwik Mukherjee\aff{1}, John D. Gibbon\aff{2} \and Dario Vincenzi\aff{3}}

\affiliation{\aff{1}International Centre for Theoretical Sciences, Tata Institute of Fundamental Research, \linebreak Bengaluru 560089, India
\aff{2}Department of Mathematics, Imperial College London SW7 2AZ, UK
\aff{3}Universit\' e C\^ote d’Azur, CNRS, LJAD, 06100 Nice, France}

\newcommand{\Gr}{\mathit{Gr}}
\newcommand{\hs}[2]{\Vert #1 \Vert_{\dot{H}^{#2}}}
\newcommand{\ltwo}[1]{\Vert #1 \Vert_{L^2}}
\newcommand{\tavg}[1]{\left\langle #1 \right\rangle}

\definecolor{dartmouthgreen}{rgb}{0.05, 0.5, 0.06}

\newcommand{\rem}[1]{}

\begin{document}
\maketitle

\begin{abstract}
In the standard theoretical setting of body-forced turbulence, the forcing that sustains the flow is concentrated in a narrow range of length scales. However, in experiments of fractal-grid turbulence and in numerical simulations inspired by the renormalization group approach, more general forcing functions have been considered. These studies have shown that the phenomenology of turbulence is sensitive to the regularity of the forcing, which raises the wider question of the sensitivity of all Navier--Stokes mathematical estimates to the regularity of body forces. To answer this question, it is necessary to convert the traditional estimates based on the Grashof number, a dimensionless measure of the magnitude of the forcing, to a form dependent on the Reynolds number, the usual dimensionless number in experimental measurements and statistical theories of turbulence. To investigate these issues we consider the two-dimensional case and employ the full range of forcing regularity allowed by the theory of weak solutions to extend available estimates not only for the energy and enstrophy dissipation rates, but also for the dimension of the global attractor. What emerges is the existence of three distinct regimes as a function of the regularity of the forcing.
\end{abstract}



\section{Introduction}
\label{sect:intro}

\subsection{The body-forced Navier-Stokes equations}

The complexity and limited predictability of turbulent flows justify the use of statistical tools to describe their dynamics. These have been successful in capturing key aspects of turbulence, offering explanations that align well with experimental observations \citep{my75,frisch1995,verma2019,bt23}. An independent theoretical route has been to consider the Navier-Stokes equations (NSEs) as a deterministic, infinite-dimensional dynamical system \citep{dg95,fmrt01,constantin2006,doering2009,bt13,robinson2016,vicol2022}.  Interestingly, the two approaches overlap in that statistical predictions often correspond to the saturation of bounds on norms of Navier–Stokes weak solutions \citep{foias_1997,doering1999,fmrt01}. However, formidable difficulties appear in any attempt to derive these bounds in physically realistic settings, such as in the presence of boundaries where vorticity can be created or destroyed in an uncontrolled manner. This is why it has been traditional to study the NSEs in periodic domains, as is also the case for many direct numerical simulations. In the absence of boundaries, a steady-state solution can be obtained if a forcing function $\bm f(\bm x)$ is added to the NSEs to form what are called the body-forced Navier-Stokes equations on the periodic domain $\Omega=[0,\ell]^d$ with $d=2,3$\,:
\begin{equation} \label{eq:NS-u}
    \partial_t \bm u + \bm u \bcdot \bnabla\bm u  
    = - \bnabla p + \nu \Delta \bm u + \bm f\,, \qquad \bnabla \bcdot \bm u = 0\,,
\end{equation}
where $\bm u$ is the velocity field, $p$ is the pressure and $\nu$ is the kinematic viscosity. It is also assumed that both $\bm f(\bm x)$ and the initial velocity $\bm u_0(\bm x)$ are divergence-free and have zero spatial mean, so that $\bm u(\bm x,t)$ remains divergence-free and mean-zero at all times. In most studies, $\bm f$ is taken time-independent. 
\par\smallskip
Mathematical estimates for the body-forced NSEs were originally expressed in terms of the dimensionless Grashof number $\Gr$, which is a measure of the magnitude of the forcing and is thus a control parameter of the system \citep*{fmtt83}. Before giving the definition of $\Gr$, let us remark that while this choice may be  natural from a mathematical point of view, experimental measurements and statistical theories of turbulence are typically expressed in terms of the Reynolds number $\Rey$, which is based on the velocity and is therefore a measure of the system response. A meaningful comparison between mathematical and physical results thus requires a conversion of bounds expressed in terms of $\Gr$ into an $\Rey$-dependent form.
\par\smallskip
With the standard notation for the $L^2$-norm and the long-time average
\begin{equation}\label{normdef}
\|\cdot\|_{L^2}^{2} = \int_{\Omega}|\cdot|^{2}\,\mathrm{d}{\bm x}\qquad\mbox{and}\qquad
\langle \cdot  \rangle = \limsup_{T\to\infty} \frac{1}{T} \int_{0}^{T} \cdot \ {\rm d}t\,,
\end{equation}
the  definitions of $\Gr$ and $\Rey$ are
\begin{equation}\label{GrRedef}
\Gr = \frac{\ell^{3 - d/2}\|\bm f\|_{L^2}}{\nu^{2}}\,,\qquad \qquad \Rey = \frac{U\ell}{\nu}\,, \qquad\qquad 
U^{2} = \ell^{-d}\left<\|\bm u\|_{L^2}^{2}\right>\,,
\end{equation}
where it is assumed that ${\bm f}\in L^2(\Omega)$. Note that $U$ is a bounded quantity for Navier-Stokes weak solutions. For periodic boundary conditions and square-integrable forcing, a conversion of bounds in terms of $\Gr$ into those in terms of $\Rey$ was first carried out by \citet{df02} in a seminal paper in which they showed that, for Navier-Stokes solutions in any dimension, 
\begin{equation}\label{DF02}
\Gr\leqslant c_1 \Rey^2 + c_2 \Rey\,.
\end{equation}
The pre-factors $c_1$ and $c_2$ depend only on the functional shape of the forcing and are uniform in the other system parameters. The formula in (\ref{DF02}) is remarkable not only for its simplicity, but also because it connects the results of analysis on the body-forced problem with the results of numerical and laboratory experiments. Moreover,  \citet{df02} also derived a rigorous upper bound for the time-averaged energy dissipation rate 
\begin{equation}\label{eq:eps-def}
	 \varepsilon = \nu \ell^{-d} \left \langle \int_{\Omega} {|\bnabla}{\bm u}|^2 {\rm d}{\bm x} \right \rangle,
\end{equation}
which takes the form
\begin{equation}
\label{eq:eps-df}
    \frac{\varepsilon\ell}{U^3}\leqslant c_1 + \frac{c_2}{\Rey}\,.
\end{equation}
This bound captures the laminar flow behaviour at small $\Rey$ and, in three dimensions, is consistent with the dissipative anomaly in the infinite-$\Rey$ limit. An estimate of $\varepsilon$ for body-forced turbulence in terms of $\Rey$ was first derived by \citet{foias_1997}\,; however, that estimate included a spurious volume-dependent pre-factor. Under the assumption of narrow-band forcing, the bound in \eqref{eq:eps-df} was later extended to moments of the velocity derivatives of any order \citep{gibbon19,gibbon2020epl}.

\subsection{Why the regularity of the forcing function is an important issue}\label{sec:why-force}

Not only is the $\Rey$-number dependence of bounds on the dissipation rates an important goal, but of equal importance is how the regularity (smoothness) of the forcing function $\bm f$ affects these bounds and ultimately the evolution of all Navier-Stokes vector fields. The results of \cite{df02} hold for $\bm f\in L^2(\Omega)$, but weak solutions of the NSEs exist for less regular forcing functions \citep{robinson2001}. It is therefore interesting to understand how the results change when $\bm f$ is not square-integrable. The regularity of $\bm f$ is more than just a technical point, because it also arises as a key issue in experiments and numerical simulations with broad-band forces, the aim of which is to investigate a potential breakdown of the classical turbulent cascade phenomenology. 
\par\smallskip
In three dimensions, Kolmogorov's theory assumes that energy injection is concentrated around a characteristic length scale, which itself is much larger than the viscous scale \citep{frisch1995}. Forcings whose action extends over a wide range of scales, and even reach the viscous range, do not display the necessary scale separation between injection and dissipation. This overlap leads them to generate a different phenomenology. Experiments on turbulent flows generated by fractal grids have clearly illustrated this point by showing a dependence of the dissipation law and the scaling properties of the turbulent field on the features of the grid (see \citet{vassilicos2015} and references therein). In numerical simulations, the action of a fractal grid has been modelled by using body forces whose Fourier spectrum scales as a power of the wavenumber $k$ \citep*{mazzi2002,mazzi2004fractal}. In this case, if the forcing spectrum is shallow enough, the law of finite energy dissipation is violated, so $\varepsilon$ grows with $\Rey$.
\par\smallskip
Power-law forcings have also been at the heart of renormalization-group (RG) approaches, whose main objectives have been the calculation of the Kolmogorov constant and the determination of the scaling of the energy spectrum \citep{frisch1995,smith1998renormalization,adzhemyan1999field,zhou2010renormalization}. The canonical setting is  white-in-time Gaussian body forces whose covariance in Fourier space scales like $k^{4-d-y}$, where $y$ is an expansion parameter. Using a body force with power-law correlation is indeed a strategy for introducing a small parameter into the NSEs, though the {\it functional} RG-formalism now allows for non-perturbative approaches \citep{canet_2022}. The same kind of forcing has been employed beyond the RG-framework to investigate the effect of injection mechanisms on the statistics of the velocity field (\citealt{sain1998,biferale2004}\,; \citealt{Biferale_2004,mitra2005stochastic}\,; \citealt*{de2023dynamic}). In particular, in three dimensions, the critical threshold $y_{c}=4$ separates two regimes\,: for $y<y_{c}$ the small scales are forcing-dominated and display nearly dimensional scaling, whereas for $y>y_{c}$ the small-scale statistics become essentially forcing-independent and exhibit the same anomalous scaling found in flows forced only at large scales \citep{Biferale_2004}. Thus, these studies demonstrate the strong impact of the forcing regularity on the dissipation of energy in three dimensions. 

\par\smallskip
The appropriate functional space in which to study broad-band body forces is the homogeneous Sobolev space $\dot{H}^s(\Omega)$. Consider a vector field $\bm g(\bm x,t)$ such that its spatial mean is zero for all $t$ and define its Fourier coefficients $\hat{\bm g}_{\bm k}(t)$ such that
\begin{equation}
\bm g(\bm x,t) = 
\sum_{\bm k} \hat{\bm g}_{\bm k}(t) e^{i\bm{k}\bcdot\bm x}, \qquad  \hat{\bm g}_{\bm k}(t) = 
\ell^{-d}\int_{\Omega} \bm{g}(\bm x,t) e^{-i\bm k\bcdot\bm x}{\rm d}{\bm x}\,.
\end{equation}
Here, all Fourier series are understood to be over the wave-vectors $\bm k =2\pi\bm n/\ell$ such that $\bm n\in\mathbb{Z}^d$ and $k=\vert\bm k\vert\neq 0$, since only mean-zero vector fields will be considered. For $\bm g(\cdot,t) \in \dot{H}^{s}(\Omega)$ we require that $\bm g(\cdot,t)$ has zero mean with the square of the norm defined as
\begin{equation}
\label{eq:Hs}
\Vert \bm g \Vert^2 _{\dot{H}^s} =
\int_\Omega \big\vert(-\Delta)^{\frac{s}{2}} \bm g(\bm x,t)\big\vert^2  \,
\mathrm{d}\bm x = \ell^d \sum_{\bm k}  k^{2s}|\hat{\bm g}_{\bm k}(t)|^2 < \infty\,.
\end{equation}%
Thus, in simple terms, $\dot{H}^{s}(\Omega)$ is the space of mean-zero fields in $d$ spatial dimensions whose Fourier coefficients decay at large $k$ faster than $k^{-s-d/2}$. The parameter $s$ characterizes the regularity of the field, with higher values of $s$ corresponding to smoother fields. In particular, the case $s=0$ corresponds to the $L^2$-norm. Homogeneous Sobolev spaces also possess a scaling property\,; i.e. for all $\lambda > 0$, the norm obeys the relation
\begin{equation}\label{Hdotdef2}
    \|\bm g (\lambda\bm x,t)\|_{\dot{H}^{s}} = \lambda^{s-d/2}
    \|\bm g (\bm x,t)\|_{\dot{H}^{s}}\,.
\end{equation} 
Weak solutions of the body-forced NSEs exist for $\bm f\in \dot H^s(\Omega)$ with $s\geqslant -1$ \citep{robinson2001}. For instance, 
at the extreme end of the range when $s=-1$, $\bm f$ is not a proper function but is more like a distribution. As noted earlier, the bound in \eqref{eq:eps-df} holds only for square-integrable body forces ($s\geqslant 0$) and does not cover the case of less regular forcing functions in the range $0 >s\geqslant -1$. In three dimensions, this case was studied by \citet*{cdp07}\,: note that their definition of the $\dot{H}^s$-norm differs by a factor of $\ell^{s-d/2}$ from the standard definition given in \eqref{eq:Hs}.  For $0 >s\geqslant -1$, the bound on $\varepsilon$ is modified as follows
\begin{equation}
    \frac{\varepsilon \ell}{U^3} \leqslant \Rey^{-\frac{s}{2+s}}\big(c_1 + c_2\,\Rey^{-1}\big)^{\frac{2}{2+s}}\,.
\end{equation}
Therefore, the value $s=0$ marks the transition between two different regimes. If $\bm f$ is square-integrable, the bound on $\varepsilon$ is independent on $s$ and is consistent with a residual finite dissipation in the infinite-$\Rey$ limit. 
For less regular $\bm f$, the dissipation law varies with the degree of regularity of the body force, and $\varepsilon$ can grow as a power of $\Rey$ with an exponent that depends on $s$. The existence of these two regimes is consistent with both the theoretical predictions and the numerical results for stochastic forcing \citep{biferale2004}. A growth of $\varepsilon$ with $\Rey$ was also observed in simulations of fractal-grid turbulence \citep{mazzi2004fractal}.

\subsection{The two-dimensional Navier-Stokes case}

In two dimensions, the NSEs possess two quadratic invariants in the sense of strong solutions\,; namely, energy and enstrophy. The respective time-averaged dissipation rates are $\varepsilon$, defined as in \eqref{eq:eps-def}, where now $d=2$, and
\begin{equation}\label{chidef}
\chi = \nu \ell^{-2} \left \langle \int_{\Omega} |\bnabla \omega|^2 {\rm d}{\bm x} \right\rangle\,,
\end{equation}
where  
\begin{equation}
\label{eq:vort}
    \omega = \bm e_3 \bcdot (\bnabla \times \bm u)
\end{equation}
is the scalar vorticity.
As in the three-dimensional case, the traditional estimates of $\varepsilon$ and $\chi$ assume that ${\bm f}\in L^2(\Omega)$ and are found in terms of  the Grashof number defined in \eqref{GrRedef} (\citealt*{dg95,foiasmanleytemam1993}; \citealt{fjmr02}; \citealt*{dfj08}) 
\begin{equation}
\label{eq:bounds-G}
    \varepsilon \leqslant \nu^3\ell^{-4}\,\Gr^{2}\,,
    \qquad
    \chi \leqslant \nu^3\ell^{-6}\,\Gr^{2}\,.
\end{equation}
While a bound for $\chi$ is not available for less regular body forces, for $\varepsilon$ to be bounded, it is sufficient that
$\bm f\in\dot{H}^{-1}(\Omega)$. In this case
\begin{equation}
\label{eq:eps_Grstar}
\varepsilon \leqslant \nu^3\ell^{-4} \Gr_{*}^{2}\,,
\end{equation}
where $\Gr_*=\nu^{-2}\ell\hs{\bm f}{-1}$ is a Grashof number based on the $\dot{H}^{-1}$-norm. If $\bm f\in L^2(\Omega)$, \eqref{eq:eps_Grstar} is an improvement on the corresponding bound  in \eqref{eq:bounds-G}, since $\Gr_*\leqslant\Gr$ \citep*{robinson2003,tsc04}. 
\par\smallskip
The above estimates are based on the magnitude of the body force, but once again the comparison with experimental and numerical results requires conversion  to a $\Rey$-dependent form. Rigorous upper bounds in terms of $\Rey$ were obtained by \cite{ad06} and \cite{gp07} for periodic boundary conditions and under the assumption that $\bm f\in\dot{H}^2(\Omega)$\,; i.e.~the second spatial derivative of the body force is square integrable. The strategy employed was to apply the methods of \cite{df02} to the vorticity equation instead of the equations for the velocity. For $\bm f\in\dot{H}^{2}(\Omega)$, the estimates equivalent to  \eqref{eq:bounds-G} are \citep{ad06,gp07}
\begin{equation}
    \frac{\varepsilon \ell}{U^3}  \leqslant \Rey^{-1/2}\left(c_1 + \frac{c_2}{\Rey}\right)^{1/2},
    \qquad
    \frac{\chi \ell^3}{U^3} \leqslant c_1 + \frac{c_2}{\Rey}\,.
\end{equation}
 The infinite-$\Rey$ limit of these bounds is consistent with the Kraichnan-Leith-Batchelor (KLB) theory, which states that an inverse transfer of energy to large scales coexists with a direct cascade of enstrophy to small scales  \citep[see][for a review]{be12,ab18}. Indeed, this phenomenology suggests that in a forced state $\varepsilon$ vanishes as viscosity tends to zero, while $\chi$ remains finite.
 \par\smallskip
 More stringent bounds apply when the forcing is monochromatic or injects energy in a finite set of wavenumbers at a constant rate. In these cases, both $\varepsilon$ and $\chi$ scale like $\Rey^{-1}$ and vanish as $\Rey$ tends to infinity\,; therefore, the enstrophy cascade disappears in this limit \citep*{cfm94,tran2002constraints,ad06}. An analogous result holds for white noise forcing \citep{eyink1996exact}. Incidentally, enstrophy dissipation also vanishes in the inviscid limit when forcing is absent and turbulence decays at long times \citep*[][see also \citealt*{mpy22} and references therein]{td06,mazzuccato06}.
\par\smallskip
Another crucial difference between two and three dimensions is that, if $\bm f\in L^2(\Omega)$, the two-dimensional NSEs on a periodic domain are known to possess a global attractor $\mathcal{A}$. The global attractor $\mathcal{A}$ is a compact set on which the long-time dynamics is confined \citep[][see also \citealt{robinson2001,robinson2013}]{lad75}. Thus, asymptotically in time, any initial condition outside $\mathcal{A}$ approaches it, while any trajectory starting on $\mathcal{A}$ remains there for all time. The existence of such an attractor reflects the effectively finite-dimensional behaviour of an otherwise infinite-dimensional dynamical system. Bounds for the fractal dimension of the attractor, $d_f(\mathcal{A}$), indeed provide a mathematically rigorous notion of the `degrees of freedom' of the system \citep[see][for an informal review]{robinson1995finite,robinson2007}. Restricting to periodic boundary conditions -- the setting considered here -- \cite*{cft88} were the first to obtain an estimate of $d_f(\mathcal{A})$ in terms of $\Gr$ (see also \citet{dg91} for a simpler proof based on the vorticity rather than the velocity field). The result is 
\begin{equation}\label{2Dattdim-G}
d_f(\mathcal{A}) \leqslant c_3 \Gr^{2/3}(1+\ln\Gr)^{1/3}\,,
\end{equation}
where $c_3$ is a positive dimensionless constant. When the forcing is at small scales, the bound in \eqref{2Dattdim-G} is improved by a bound that is linear in $\Gr_*$ \citep{robinson2003,tsc04}.  Later,  \citet{gp07} converted \eqref{2Dattdim-G} into a $\Rey$-dependent estimate under the more restrictive assumption of $\dot{H}^2$-forcing
\begin{equation}\label{attdim-Re}
d_f(\mathcal{A}) \leqslant c_4\Rey(1+\ln \Rey)^{1/3}\,,
\end{equation}
where $c_4$ is another positive dimensionless constant. 
The $\Rey$-based estimate in \eqref{attdim-Re} matches the heuristic prediction for the degrees of freedom of two-dimensional turbulence. Indeed, if $\eta_\chi$ denotes the smallest length scale in the system, the number of degrees of freedom of the system can be related to the number of boxes of size $\eta_\chi$ that fit into the domain $\Omega$, which leads to the identification $d_f(\mathcal{A})\sim (\ell/\eta_\chi)^{2}$. 
Comparison of this interpretation of $d_f(\mathcal{A})$ with \eqref{attdim-Re} yields an estimate of the enstrophy dissipation scale
\begin{equation}\label{attdim-eta}
\eta_\chi\sim \ell\Rey^{-1/2}(1+\ln \Rey)^{-1/6}\,,
\end{equation}
which is consistent with the KLB theory \citep{be12} to within logarithmic corrections. 

\begin{figure}
    \centering
    \includegraphics[width=1.0\linewidth]{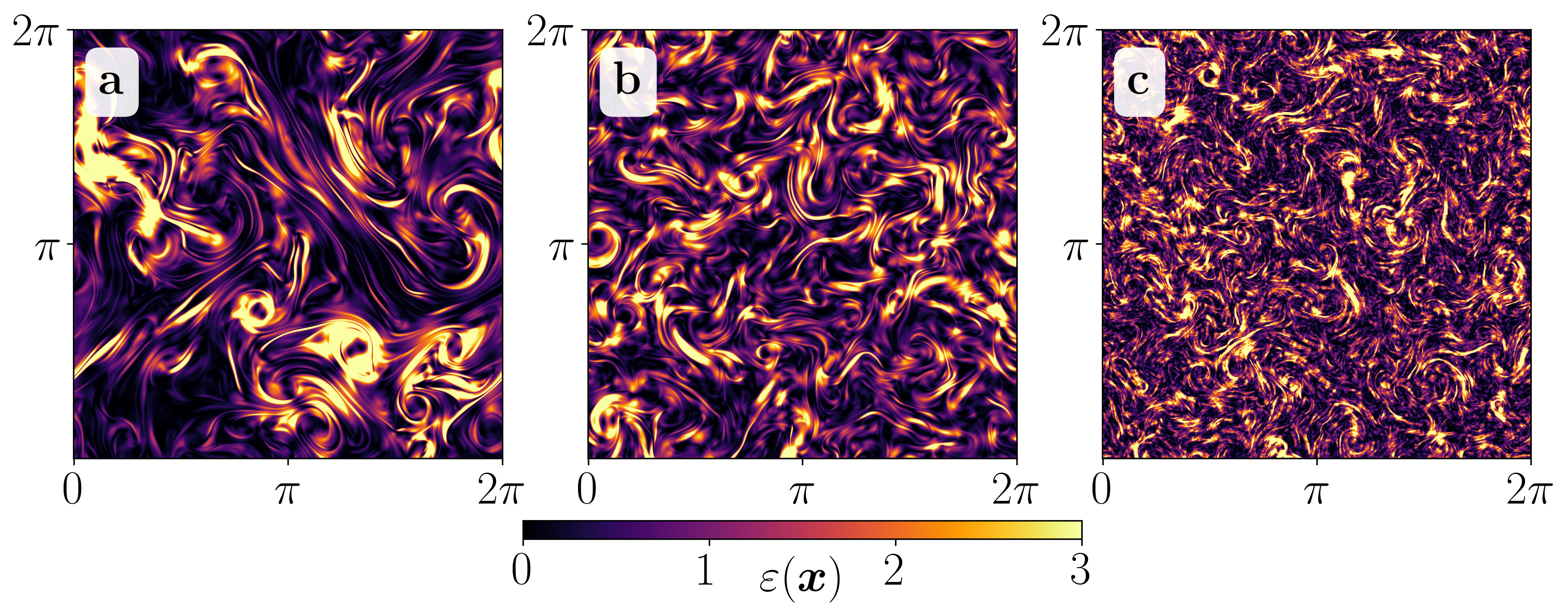}
    \includegraphics[width=1.0\linewidth]{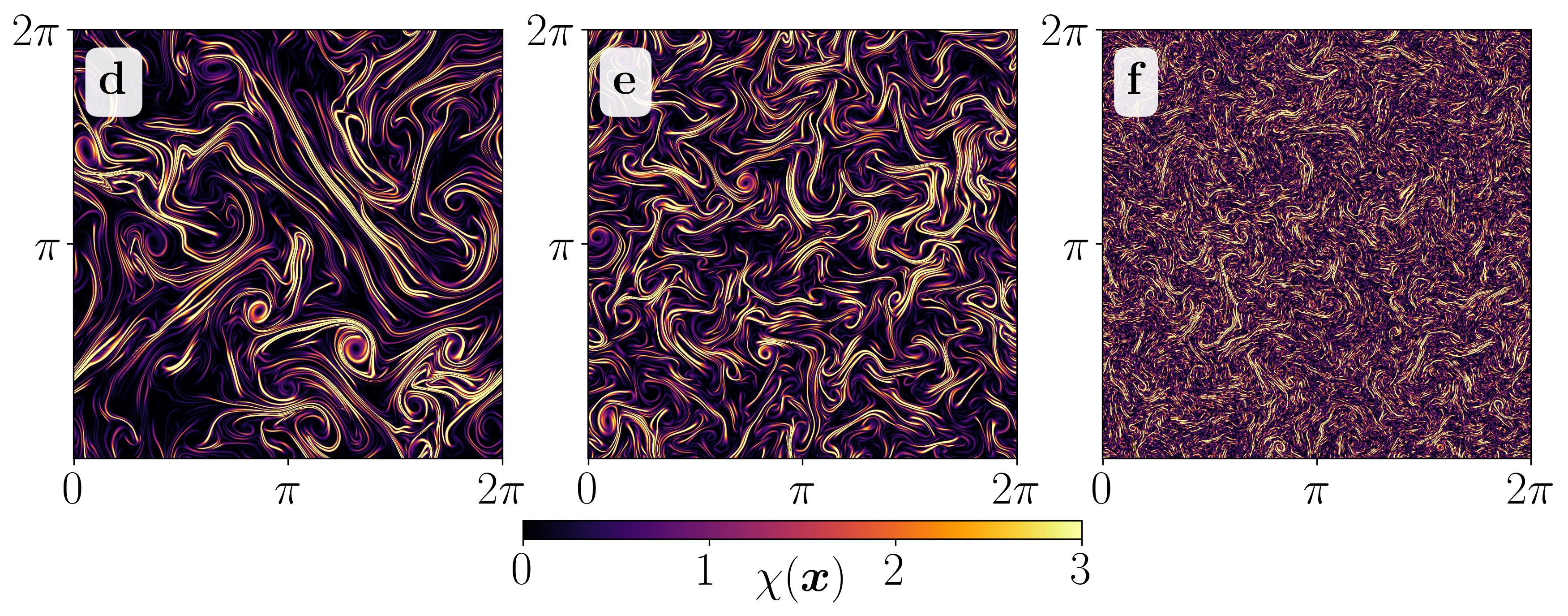}
    \caption{The panels show the instantaneous energy and enstrophy dissipation fields for two-dimensional Navier-Stokes turbulence, defined as $\varepsilon(\bm x) = 2\nu \operatorname{Tr}(\boldsymbol{\mathsf{S}}^2)$, with $\boldsymbol{\mathsf{S}}$ being the rate-of-strain tensor,  and $\chi(\bm x)=\nu|{\bm \nabla}\omega|^2$, respectively. The data are from a pseudospectral simulation in a $2\pi$-periodic domain with $1024^2$ collocation points, shown at a representative time in the statistically steady state. Large-scale friction has been added to stabilize the system. Panels (\textit{a}--\textit{c}) show the energy dissipation fields for three different forcings, while the corresponding enstrophy dissipation fields are given in panels (\textit{d}--\textit{f}) for the same three forcing configurations\,: (\textit{a},\,\textit{d}) monochromatic forcing with wavenumber $k_f = 10$\,; (\textit{b},\,\textit{e}) power-law forcing such that $\vert\hat{\bm f}_{\bm k}\vert\propto k^{-4}$ for $k \in [10, 256]$ and zero otherwise (this forcing mimics the case $\bm f\in \dot{H}^s(\Omega)$ with $s \geqslant  2$)\,;
    (\textit{c},\,\textit{f}) the same as in (\textit{b},\,\textit{e}) but with $\vert\hat{\bm f}_{\bm k}\vert\propto k^{-2}$ (this forcing mimics the case $\bm f\in \dot{H}^s(\Omega)$ with $0< s < 2$). Panels (\textit{a}) and (\textit{b}), and equivalently panels (\textit{d}) and (\textit{e}), are qualitatively similar\,: for $s \geqslant 2$ the forcing spectrum decays steeply, so few modes beyond $k_f = 10$ are effectively forced, and only slight differences are  visible at small scales. By contrast, for $0<s<2$ the forcing spectrum decays slowly enough to excite a wide range of modes which generates dissipation fields with a substantially different small-scale structure.}\label{MGVfig}
\end{figure}

\subsection{The goal of this study}

In two dimensions we have seen that the available $\Rey$-dependent estimates of $\varepsilon$, $\chi$ and $d_f(\mathcal{A})$ require that the body force possesses a relatively high degree of regularity. Indeed, these estimates assume that $\bm f\in \dot{H}^2(\Omega)$ which, in two dimensions, means that $\vert\hat{\bm f}_{\bm k}\vert^2$ must decay faster than $k^{-6}$. However, weak solutions of the NSEs exist for more general forcing functions, namely for $\bm f\in \dot{H}^{-1}(\Omega)$, i.e. for any $\bm f$ such that $\vert\hat{\bm f}_{\bm k}\vert^{2}$ grows like $k$. Moreover, analytical predictions and numerical simulations for stochastic forcing have shown that, even in two dimensions, the regularity of the body force has a strong impact on the turbulent flow \citep*{mazzino2007PRL,mazzino2009IOP}. In particular, these studies have shown the existence of different dynamical regimes, the KLB scenario being observed only when the forcing is sufficiently smooth. The panels in figure~\ref{MGVfig}, which are included for illustrative purposes, display contrasting results for the instantaneous energy dissipation field, and the enstrophy dissipation field, when different forcings are used. Even in two dimensions, the question therefore arises as to the effect of the forcing regularity on Navier--Stokes mathematical estimates. Unlike the three-dimensional case, to the best of our knowledge, this question has not yet been addressed. Therefore, our aim here is to extend the existing $\Rey$-dependent estimates for $\varepsilon$ and $\chi$ to body forces in $\dot{H}^s(\Omega)$ with $-1 \leqslant s < 2$. Moreover, we also extend the estimate of $d_f(\mathcal{A})$ to the range $0\leqslant s < 2$, where a global attractor is known to exist. To this end, we combine the methods of \cite{ad06} and \cite{gp07} with those of \cite{cdp07}. In essence, our strategy is to apply the techniques developed for the three-dimensional NSEs with $\dot{H}^s$-forcing  to the vorticity equation in two dimensions.
\par\smallskip
Our results for $\varepsilon$, $\chi$ and $d_f(\mathcal{A})$ are summarized in table~\ref{table}, in which only the large-$\Rey$ bounds are shown. While the first line represents the special case of monochromatic or constant-$\varepsilon$ forcing, the rest of the table refers to the case of $\dot{H}^s$-forcing. Our analysis reveals the existence of three distinct regimes depending on the value of $s$. The case $s\geqslant 2$ is that studied by \citet{ad06} and \citet{gp07}. Our estimates correspond to $0\leqslant s< 2$. The last line ($-1\leqslant s< 0$) is a straightforward adaptation of the three-dimensional result of \citet{cdp07} to two dimensions. The rest of this article is organized as follows. In \S\ref{sect:derivs} we present the derivation of our estimates for constant-in-time forcing. In \S\ref{sect:time-dependent} we extend the results to time-dependent body forces whereas in \S\ref{sect:conclusions} we summarize the results associated with the various dynamical regimes which are a function of the regularity of the forcing and discuss their main implications. 
\begin{table}
	\begin{center}
	\def~{\hphantom{0}}
    \begin{tabular}{l|c|c|c|c}
		$\dot{H}^s$ forcing & $\chi \ell^3/U^3$ & $ \varepsilon \ell/U^3$ & $\ell\eta_\chi^{-1}$ 
        & $d_f(\mathcal{A})$ 
		\\[7pt]\hline
		single-$k$ or fixed-$\varepsilon$ & $\mathit{Re}^{ -1}$ & $\mathit{Re}^{ -1}$ & $\mathit{Re}^{ 1/3}$ 
        & $\mathit{Re}^{ 2/3}(1+\ln \mathit{Re})^{ 1/3}$		
		\\[3pt]
		$s\geqslant 2$ & $\mathit{Re}^{ 0}$ & $\mathit{Re}^{-1/2}$ & $\mathit{Re}^{1/2}$
        & $\mathit{Re}(1+\ln \mathit{Re})^{ 1/3}$
		\\[3pt] 
		$0\leqslant s < 2$ & $\quad \mathit{Re}^{ \tfrac{2-s}{2+s}}$ \quad & $ \quad \mathit{Re}^{ -\tfrac{s}{s+2}} \quad $ & $\quad \mathit{Re}^{\tfrac{4+s}{3(2+s)}} \quad $ 
        & $\quad \mathit{Re}^{\tfrac{2(4+s)}{3(2+s)}}\left(1+\ln \mathit{Re} \right)^{ 1/3} \quad $  \\[3pt]
		$-1 \leqslant s < 0$ & --- & $\mathit{Re}^{-\tfrac{s}{s+2}}$ & ---
        &--- 
	\end{tabular}
\caption{Large-$\Rey$ estimates for the two-dimensional NSEs with $\dot{H}^s$ constant-in-time forcing. If the forcing is time dependent, the estimates for the time-averaged dissipation rates are multiplied by a pre-factor that depends on the Strouhal number, i.e.~the ratio of the convective time of the flow to the characteristic time scale of the forcing (see \S\ref{sect:time-dependent} for the details).}\label{table}
\end{center}
\end{table}

\section{Estimates for $\dot{H}^s$ forcing}\label{sect:derivs}

We consider the incompressible NSEs on the periodic domain $\Omega=[0,\ell]^2$ as in \eqref{eq:NS-u} with $d=2$.  To take advantage of the structure of the NSEs in two dimensions, it is convenient to study the evolution of the scalar vorticity, defined in \eqref{eq:vort}. Taking the curl of \eqref{eq:NS-u} yields the vorticity equation
\begin{equation}
	\partial_t \omega + \bm u \bcdot \bnabla \omega
	= \nu \Delta \omega + \phi\,,
	\label{eq:NS-vort}
\end{equation}
where $\phi = \bm e_3 \bcdot (\bnabla \times \bm f)$. In this section, we restrict to a constant-in-time body force ; in \S\ref{sect:time-dependent}, we shall extend the results to the case of time-dependent forcing. Here, the body force $\bm f$ is taken to be in $\dot{H}^s(\Omega)$ and of the form
\begin{equation}
    \bm f(\bm x) = F\, \bm \Phi(\ell^{-1}\bm x)\,,
\end{equation}
where $F>0$ is the magnitude and $\bm \Phi$ is a dimensionless field such that  
\begin{equation}\label{eq:shape}
\int_{\Omega} \bm \Phi(\bm x)\,{\rm d}\bm x = 0\,,
\qquad
\bnabla\bcdot\bm\Phi=0\,,
\qquad \ell^{s-1}|\|\bm \Phi\|_{\dot{H}^s} = 1\,.  
\end{equation}
We shall refer to $\bm\Phi$ as the `shape' of the body force. Recall that weak solutions of \eqref{eq:NS-u} are known to exist for $s\geqslant -1$ \citep{robinson2001,robinson2013}. Finally, it is useful to note that $\varepsilon$ and $\chi$ are expressed in terms of the norms of $\bm u$ and $\omega$ as 
\begin{equation}
\varepsilon = \nu\ell^{-2}
\tavg{\hs{\bm u}{1}^2} = \nu\ell^{-2}\tavg{\ltwo{\omega}^2}, \qquad
\chi = \nu\ell^{-2}\tavg{\hs{\omega}{1}^2} = \nu\ell^{-2}\tavg{\hs{\bm u}{2}^2}\,.
\end{equation}

\subsection{The case $0\leqslant s\leqslant 2$}\label{sect:0s2}

We first consider the range $0\leqslant s \leqslant 2$, for which \eqref{eq:NS-u} has strong solutions \citep{robinson2001,robinson2013}. The first step is to relate $\chi$ to the rate of vorticity injection by the body force. To this end, we obtain the enstrophy balance equation by multiplying \eqref{eq:NS-vort} by $\omega$ and integrating over space
\begin{equation}
\label{eq:enstrophy_balance}
    \frac{1}{2}\frac{\rm d}{{\rm d}t}\|\omega\|^2_{L^2} = - \nu\| { \bnabla}\omega\|^2_{L^2} + \int_{\Omega} \omega \phi \, \text{d}{\bm x}\,. 
\end{equation}
An application of the interpolation inequality
\begin{equation}
\label{eq:interp-balance}
\int_{\Omega} \omega\phi \, {\rm d}{\bm x} \leqslant \| \bnabla\omega\|_{L^2} \| \phi\|_{\dot H^{-1}} = \| \bnabla\omega\|_{L^2} \| \bm f\|_{L^2} 
\end{equation}
and then Young's inequality to the right hand side of \eqref{eq:interp-balance} gives
\begin{equation}\label{eq:Young}
\frac{\rm d}{{\rm d}t}\|\omega \|^2_{L^2} \leqslant -\nu\|\bnabla{\omega}\|^2_{L^2} + \nu^{-1}\|{\bm  f}\|^2_{L^2}\,.
\end{equation}
Using the Poincar\'{e} inequality on the dissipation term implies
\begin{equation}\label{eq:Poincare}
    \frac{\rm d}{{\rm d}t}\|\omega\|^2_{L^2}\leqslant -4\pi^2\nu\ell^{-2}\|{\omega}\|^2_{L^2} + \nu^{-1}\|{\bm  f}\|^2_{L^2}\,.
\end{equation}
Finally, Gronwall's inequality gives
\begin{equation}
    \|{\omega}\|^2_{L^2} \leqslant\|{  \omega}_0\|^2_{L^2}\exp(-4\pi^2\nu t/\ell^2) + \frac{\ell^2}{4\pi^2\nu^2}\|{\bm  f}\|^2_{L^2}\left[1-\exp\left(-4\pi^2\nu t/\ell^2\right)\right]\,,
\label{eq:bound_omega}
\end{equation}
where $\omega_0={\bm e_3}\bcdot (\bnabla\times {\bm u_0})$ is the initial vorticity. Therefore, $\|\omega\|^2_{L^2}$ is uniformly bounded in time, and the long-time average of its time derivative vanishes. It follows from \eqref{eq:enstrophy_balance} that the time-averaged enstrophy dissipation rate satisfies
\begin{equation}
\label{eq:omega-f}
		\chi = \ell^{-2}\left \langle  \int_{\Omega}\omega  \phi \, {\rm d}{\bm x} \right \rangle\,.
\end{equation}
To estimate the right-hand side, we first recall that the Fourier transform of the curl of a divergence-free field $\bm g(\bm x,t)$ satisfies the identity 
\begin{equation}\label{eq:curl-k}
\vert (\widehat{\bnabla\times \bm g})_{\bm k}\vert^2 = \vert \bm k\times\bm \hat{\bm g}_{\bm k}\vert^2 = k^2\vert\hat{\bm g}_{\bm k}\vert^2 - (\bm k\bcdot \hat{\bm g}_{\bm k})^2=k^2\vert\hat{\bm g}_{\bm k}\vert^{2}\,. 
\end{equation}
We then use Parseval's theorem, the triangle inequality, and identity \eqref{eq:curl-k} to obtain
\begin{align}
	\ell^{-2}\int_{\Omega}\omega \phi \, {\rm d}{\bm x}  
     &\leqslant \ell^{-2}\left\vert\int_{\Omega} (\bnabla\times\bm u)\bcdot(\bnabla\times\bm f)\,{\rm d}{\bm x}  \right\vert\nonumber\\
&\leqslant
     \sum_{\bm k}  \left\vert\bm k\times \hat{\bm u}_{\bm k}\right\vert\,
     \vert\bm k\times \hat{\bm f}_{\bm k}\vert
     = \sum_{\bm k} k^2 \vert \hat{\bm u}_{\bm k}\vert
     \vert \hat{\bm f}_{\bm k}\vert\,.
\end{align}
Rearranging the sum and applying the Cauchy--Schwarz inequality gives
\begin{align}
	\ell^{-2} \int_{\Omega}\omega \phi \, {\rm d}{\bm x} &\leqslant  \sum_{\bm k} k^{2} |\hat {\bm u}_{\bm k} | |\hat {\bm f}_{\bm k}|	
	= \sum_{\bm k}  \left(k^{2-s} |\hat {\bm u}_{\bm k}|\right)  \left(k^{s}|\hat{\bm f}_{\bm k}|\right) \nonumber \\
	&\leqslant \bigg[ \sum_{\bm k}  k^{2(2-s)} |\hat{\bm u}_{\bm k}|^2 \bigg]^{1/2} \bigg[ \sum_{\bm k}  k^{2s} |\hat{\bm f}_{\bm k}|^2 \bigg]^{1/2}\nonumber\\
	&=  \ell^{-s-1} F \Vert \bm u\Vert _{\dot{H}^{2-s}}\,. 
	\label{eq:u_2_s}
\end{align}	
To relate $\chi$ to $\Rey$, we need to estimate the time average of $\Vert \bm u\Vert_{\dot{H}^{2-s}}$ in terms of $\chi$ and $U$. 
The following interpolation identity is obtained by rearranging the terms in $\Vert \bm u\Vert_{\dot{H}^{2-s}}$ and then using a H\"older inequality
\begin{align}
	\Vert \bm u\Vert^2 _{\dot{H}^{2-s}} 
    &= \ell^{2}\sum_{\bm k}  k^{2(2-s)} |\hat{\bm u}_{\bm k}|^2 
	= \ell{^2}\sum_{\bm k} \big( k^4|\hat{\bm u}_{\bm k}|^2\big)^{\frac{2-s}{2}} 
		\, \big(|\hat{\bm u}_{\bm k}|^2 \big)^{1-\frac{2-s}{2}}\nonumber \\
	&\leqslant \ell^{2}\bigg(\sum_{\bm k}  k^4 |\hat{\bm u}_{\bm k}|^2\bigg)^{1-\frac{s}{2}} 
	\bigg( \sum_{\bm k} |\hat{\bm u}_{\bm k}|^2 \bigg)^{\frac{s}{2}}\nonumber\\
	&=
    \Vert \bm u\Vert _{\dot{H}^2}^{2-s} \Vert \bm u\Vert _{L^2}^{s}\,,
	\label{u_h_s_2_final}
\end{align}
where we have used the fact that $0\leqslant s\leqslant 2$. Taking the square root of both sides, time averaging, and using a H\"older inequality leads to
\begin{equation}
	\langle \Vert \bm u\Vert _{\dot{H}^{2-s}} \rangle  \leqslant \Big \langle \Vert \bm u\Vert _{\dot{H}^2}^{1-\frac{s}{2}} \Vert \bm u\Vert _{L^2}^{\frac{s}{2}} \Big\rangle 
	\leqslant \Big\langle   \Vert \bm u\Vert _{\dot{H}^2}^2 \Big\rangle^{\frac{2-s}{4}} \Big\langle \Vert  \bm u\Vert_{L^2}^{\frac{2s}{2+s}} \Big\rangle^{\frac{2+s}{4}}\,.
\end{equation}
An application of Jensen's inequality to the last term, together with the definitions of $\chi$ and $U$, gives
\begin{align}
   \langle \Vert \bm u\Vert _{\dot{H}^{2-s}} 
   \rangle 
   &\leqslant
\Big\langle \Vert\bm u \Vert _{\dot{H}^2}^2\Big\rangle^{\frac{2-s}{4}} \Big\langle  \Vert \bm u\Vert _{L^2}^2\Big\rangle ^{\frac{s}{4}}\nonumber\\
   &=
   \nu^{\frac{s-2}{4}}\ell\,
   \Big\langle  \nu\ell^{-2}\Vert \bm u \Vert _{\dot{H}^2}^2\Big\rangle^{\frac{2-s}{4}} \Big\langle\ell^{-2} 
   \Vert \bm u\Vert _{L^2}^2\Big\rangle ^{\frac{s}{4}}\nonumber\\
   &=
   \nu^{\frac{s-2}{4}}\ell\,\chi^{\frac{2-s}{4}} U^{\frac{s}{2}}\,.
   \label{eq:chi_3}
\end{align}
By combining \eqref{eq:omega-f}, \eqref{eq:u_2_s} and \eqref{eq:chi_3}, we find
\begin{eqnarray} 
	\chi \leqslant  F  \nu^{\frac{s-2}{4}} \ell^{-s}\chi ^{\frac{2-s}{4}}U^{\frac{s}{2}},
\end{eqnarray}
which leads to
\begin{equation}
\chi \leqslant \nu^{-\frac{2-s}{2+s}} \ell^{-\frac{4s}{2+s}} F^{\frac{4}{2+s}}  U^{\frac{2s}{2+s}}\,.
\label{eq:chi_inter}
\end{equation}
It remains to estimate $F$ in terms of $U$. We take a smooth, time-independent, dimensionless field ${\bm \psi}$ such that 
\begin{equation}
\label{eq:psi}
\bnabla\bcdot\bm\psi = 0, \qquad \int_{\Omega} {\bm \Phi \bcdot {\bm \psi}} \, {\rm d}{\bm x} > 0 
\end{equation}
(${\bm \psi}$ can be, for example, a suitable Galerkin truncation of ${\bm \Phi}$). We then multiply the NSEs \eqref{eq:NS-u} by ${\bm\psi}$ and integrate over space
\begin{equation}
    \int_\Omega {\bm \psi} \bcdot  \partial_t{\bm  u}\,{\rm d}{\bm x}
    + \int_\Omega {\bm\psi} \bcdot ({\bm  u}\bcdot\bnabla){\bm  u}\,{\rm d}{\bm x} 
    = -\int_\Omega {\bm \psi} \bcdot \bnabla p\,\,{\rm d}{\bm x}
    + \nu\int_\Omega {\bm\psi} \bcdot \Delta{\bm  u}\,{\rm d}{\bm x} 
    + \int_\Omega   {\bm  f} \bcdot {\bm \psi}\,{\rm d}{\bm x}.
\end{equation}
By integrating by parts and using 
 $\bnabla\bcdot{\bm  u}=0$ on the nonlinear term and $\bnabla\bcdot\boldsymbol{\psi}=0$ on the pressure term, we find
\begin{equation}
    \int_\Omega {\bm  f} \bcdot {\bm  \psi}\,{\rm d}{\bm x} 
    = \int_\Omega {\bm  \psi} \bcdot \partial_t{\bm  u}\, {\rm d}{\bm x} 
    - \int_\Omega {\bm  u}\bcdot\bnabla{\bm  \psi}\bcdot{\bm  u}\,{\rm d}{\bm x} 
    - \nu\int_\Omega {\bm  u}\bcdot\Delta{\bm  \psi}\,{\rm d}{\bm x}.
\end{equation}
The long-time average of both sides eliminates the time-derivative term, since $\|{\bm  u}\|_{L^2}^2$ is uniformly bounded in time \citep[see][]{df02}. Therefore
\begin{equation}
    \int_\Omega {\bm f}\bcdot{\bm\psi}\,{\rm d}{\bm x} 
    \leqslant \left|\left\langle\int_\Omega {\bm u}\bcdot\bnabla{\bm\psi}\bcdot{\bm u}\,{\rm d}{\bm x}\right\rangle\right| 
    + \nu\,\left|\left\langle\int_\Omega {\bm u}\bcdot\Delta{\bm\psi}\,{\rm d}{\bm x}\right\rangle\right|.
    \label{eq:tavf}
\end{equation}
Now, we apply the Cauchy--Schwarz and Jensen's inequalities and get%
\begin{align}
 \int_{\Omega}\bm f \bcdot \bm \psi \, {\rm d}{\bm x} 	 	 &\leqslant \Vert \bnabla \bm \psi\Vert _{L^{\infty}} \left\langle \Vert \bm u\Vert _{L^2}^2 \right\rangle + \nu\, \Vert \Delta \bm \psi \Vert_{L^2}  \left\langle \Vert \bm u\Vert _{L^2} \right\rangle\nonumber \\[0mm]
         &\leqslant \Vert \bnabla \bm \psi\Vert _{L^{\infty}} \left\langle \Vert \bm u\Vert _{L^2}^2 \right\rangle + \nu\, \Vert \Delta \bm \psi \Vert_{L^2}  \left\langle \Vert \bm u\Vert _{L^2}^2 \right\rangle^{1/2}\nonumber \\[2mm]
	&= \ell\,\Vert \widetilde{\bnabla} \bm \psi\Vert _{L^{\infty}} U^2 + \nu\ell^{-1}  \Vert \widetilde\Delta \bm \psi \Vert_{L^2} \,  U\,,
    \label{eq:bound-F}
\end{align}
where $\widetilde{\bnabla} = \ell \bnabla$ and $\widetilde{\Delta} = \ell^2 \Delta$. After rewriting the left-hand side of \eqref{eq:bound-F} as
\begin{equation}
\int_{\Omega}\bm f \bcdot \bm \psi \, {\rm d}{\bm x} 
	 = F \int_{\Omega} {\bm \Phi} \bcdot {\bm \psi} \, {\rm d}{\bm x}\,,
     \label{eq:rhs-F}
\end{equation}
we find
\begin{equation}
F \leqslant c_{1}\ell^{-1}U^{2} + c_{2}\nu\ell^{-2} U\,,
\label{eq:F_amp}
\end{equation}
together with
\begin{equation}
	c_1 = \left(\ell^{-2}\int_{\Omega} {\bm \Phi} \bcdot {\bm \psi} \, {\rm d}{\bm x}\right)^{-1} \Vert {\widetilde{\bnabla}{\bm \psi}}\Vert_{L^{\infty}}, \qquad c_2 = \left(\ell^{-2}\int_{\Omega} {\bm \Phi} \bcdot {\bm \psi} \, {\rm d}{\bm x}\right)^{-1} 
    \ell^{-1}\Vert \widetilde\Delta \bm \psi \Vert_{L^2} \,,
\end{equation}
where $c_1$ and $c_2$ are dimensionless constants not related to the same coefficients in \S\ref{sect:intro}. Finally, combining \eqref{eq:chi_inter} and \eqref{eq:F_amp}, we obtain the estimate for the rescaled time-averaged enstrophy dissipation rate
	\begin{equation}
		\frac{\chi \ell^3}{U^3} 
		\leqslant \mathit{Re}^{\frac{2-s}{2+s}}\left( c_1 + c_2 \mathit{Re}^{-1}\right)^{\frac{4}{2+s}}\,.
		\label{eq:bound_chi}
\end{equation}
Note that the pre-factors  only depend on the shape of the forcing. For $\Rey\gg 1$, \eqref{eq:bound_chi} yields the following estimate for  the enstrophy dissipation length scale 
\begin{equation}
	\ell\eta_{\chi}^{-1} = \ell\left(\frac{\chi}{\nu^3}\right)^{1/6}
    \leqslant
	c_1' \mathit{Re}^{\frac{4+s}{3(2+s)}}\,,
    \label{enstrophy_length}
\end{equation}
with $c_1'= c_1^{2/3(2+s)}$. To estimate the time-averaged energy dissipation rate, we use the interpolation inequality in 
\eqref{u_h_s_2_final} for $s=1$
\begin{equation}
	\Vert {\bm u} \Vert_{\dot{H}^1} \leqslant \Vert  \bm u\Vert _{\dot{H}^2}^{1/2} \Vert \bm u\Vert _{L^2}^{1/2}\,. 
\end{equation}
We now square both members, time average, and use the Cauchy--Schwarz inequality to obtain
\begin{equation}
	\left\langle\Vert {\bm u} \Vert^2_{\dot{H}^1}\right\rangle
    \leqslant \left\langle \Vert  \bm u\Vert _{\dot{H}^2} \Vert \bm u\Vert _{L^2}\right\rangle 
    \leqslant \left\langle\Vert {\bm u} \Vert^2_{\dot{H}^2}\right\rangle^{1/2}
    \left\langle\Vert {\bm u} \Vert^2_{L^2}\right\rangle^{1/2}\,.
\end{equation}
After multiplying by $\nu\ell^{-2}$, we find 
\begin{equation}
    \varepsilon \leqslant \nu^{1/2} \chi^{1/2} U\,.
    \label{eq:epsilon-chi}
\end{equation}
The estimate for $\chi$ in \eqref{eq:bound_chi} can then be used to bound the right-hand side
\begin{equation}
	\frac{ \varepsilon \ell }{U^3}  \leqslant \mathit{Re}^{-\frac{s}{2+s}}\left(c_1 + c_2 \mathit{Re}^{-1} \right)^{\frac{2}{2+s}}\,.
    \label{eq:bound-eps}
\end{equation}
We conclude this study of the case $0\leqslant s\leqslant 2$ with the estimate of the attractor dimension $d_f(\mathcal{A})$ for $\Rey\gg 1$.
We recall that $d_f(\mathcal{A})$ is related to $\eta_\chi$ via the following bound \citep{cft88}
\begin{equation}
	d_{f}(\mathcal{A}) \leqslant   c_3 \big(\ell \eta_{\chi}^{-1}\big)^2 \big[1+\ln \big(\ell \eta_\chi^{-1}\big)\big]^{1/3}\,.
	\label{attractor_dim_eta}
\end{equation}
Plugging~\eqref{enstrophy_length} into~\eqref{attractor_dim_eta} gives
\begin{equation}
	d_f(\mathcal{A}) \leqslant c_3' \mathit{Re}^{\frac{2(4+s)}{3(2+s)}}\left(1+\ln \mathit{Re}^{\frac{4+s}{3(2+s)}}\right)^{1/3}
	\leqslant c_3' \mathit{Re}^{\frac{2(4+s)}{3(2+s)}}\left(1+\ln \mathit{Re} \right)^{1/3}\,,
\end{equation}
where $c'_3$ is a constant related to $c_1'$ and $c_3$. The scaling exponent varies between unity for $s=2$ and $4/3$ for $s=0$.

\subsection{The case $-1\leqslant s\leqslant 0$}\label{sec:s<0}

If $-1\leqslant s \leqslant 0$, the derivation above fails because the steps taken in \eqref{u_h_s_2_final} are invalid, so a bound for $\chi$ is not available. In particular, the absence of a bound for $\chi$ means that it is not known whether a global attractor exists in this case. However, we can obtain a bound for $\varepsilon$ by  following the same procedure used by \cite{cdp07} for  three dimensions.   This procedure uses the velocity equation instead of the vorticity equation. For completeness, we recall the main steps below (see \citealt{cdp07} for the details). Since $\Vert{\bm u}\Vert_{L^2}^2$ is uniformly bounded in time, $\varepsilon$ can be expressed in terms of the energy injection as \citep{df02,cdp07}
\begin{equation}
\varepsilon = \ell^{-2}\left\langle \int_\Omega {\bm f} \bcdot {\bm u} \,\mathrm{d}\bm x \right\rangle.
\label{eq:u-f}
\end{equation}
The integral in the right-hand side can be estimated as
\begin{align}
\ell^{-2}\int_{\Omega}{\bm f}\bcdot{\bm u}\,{\rm d}{\bm x} 
    &\leqslant \sum_{{\bm k}}|\hat{{\bm f}}_{\bm k}||\hat{{\bm u}}_{\bm k}| 
    \leqslant \bigg[\sum_{{\bm k}}k^{2s}|\hat{{\bm f}}_{\bm k}|^2\bigg]^{1/2}
         \bigg[\sum_{{\bm k}} k^{-2s}|\hat{{\bm u}}_{\bm k}|^2\bigg]^{1/2}
         \\[1mm]
         &= F\ell^{-s-1}\|{\bm u}\|_{\dot{H}^{-s}}
         \leqslant F\ell^{-s-1}\|{\bm u}\|_{\dot H^1}^{-s}\|{\bm u}\|^{1+s}_{L^2},
\end{align}
where an interpolation inequality analogous to~\eqref{u_h_s_2_final} has been used in the last step.
Taking the time average of both sides and using H\"older's and Jensen's inequalities gives
\begin{equation}
    \varepsilon 
    \leqslant F\ell^{-s-1}\big\langle\|{\bm u}\|_{\dot H^1}^{-s}\|{\bm u}\|^{1+s}_{L^2}\big\rangle
    \leqslant F\ell^{-s-1}\big\langle\|{\bm u}\|_{\dot H^1}^{2}\big\rangle^{-\frac{s}{2}}
    \Big\langle\|{\bm u}\|^\frac{2(1+s)}{(2+s)}_{L^2}\Big\rangle^\frac{2+s}{2}
    \leqslant F \,\nu^\frac{s}{2} \ell^{-s}\varepsilon^{-\frac{s}{2}}U^{1+s}.    \label{eq:bound-cheskidov}
    \end{equation}
Using the bound on $F$ in~\eqref{eq:F_amp} and rearranging gives \citep{cdp07}
\begin{equation}
    \frac{\varepsilon \ell}{U^3} \leqslant \Rey^{-\frac{s}{2+s}}\big(c_1 + c_2\,\Rey^{-1}\big)^{\frac{2}{2+s}}.
\end{equation}
The final estimate has the same form as \eqref{eq:bound-eps}, but now $-1\leqslant s<0$.

\section{Time-dependent forcing}\label{sect:time-dependent}

The estimates derived in the previous sections can be generalized to the case of deterministic but time-dependent forcing by applying the methods outlined in  \citet{ad06}. However, some variations are necessary, since here $\bm f \notin \dot H^2(\Omega)$. We give the main steps below.
\par\smallskip
The body force is now taken in the form 
\begin{equation}
\bm f(\bm x,t) = F(t) \bm\Phi(\ell^{-1}\bm x,t)\,,
\end{equation}
where $\bm\Phi$ satisfies the same properties as in \eqref{eq:shape} for all $t$. It is also assumed that the magnitude of the forcing is uniformly bounded in time and has finite mean square 
\begin{equation}
\bar{F}^2 = \big\langle F^2(t)\big\rangle < \infty\,.
\end{equation}
Let us first consider the case $0\leqslant s\leqslant 2$.
The bound  in \eqref{eq:bound_omega} continues to hold provided that $\Vert\bm f\Vert_{L^{2}}^2$ is replaced by $\sup_t\Vert\bm f\Vert_{L^{2}}^2$. Hence,
$\Vert\omega\Vert_{L^2}$ remains bounded for all $t$, and $\chi$  can again be estimated as in \eqref{eq:omega-f}.
\par\smallskip
An intermediate bound for $\chi$ 
is obtained by using the
Cauchy--Schwarz inequality to split the time average of the product of $F(t)$ and the $\dot{H}^{2-s}$-norm of the velocity and then by proceeding as in \S\ref{sect:0s2}\,:
\begin{align}
    \chi &\leqslant
     \ell^{-s-1} \left\langle F \Vert \bm u\Vert _{\dot{H}^{2-s}}\right\rangle
     \leqslant \ell^{-s-1} \bar{F} \left\langle  \Vert \bm u\Vert^2_{\dot{H}^{2-s}}\right\rangle^ {1/2}
     \leqslant \ell^{-s-1}\bar{F}\,\left\langle\|{\bm u}\|_{\dot H^2}^{2-s}\|{\bm u}\|^{s}_{L^2}\right\rangle^{1/2}
    \nonumber\\
&\leqslant \ell^{-s-1}\bar{F}\,\left\langle\|{\bm u}\|_{\dot H^2}^{2}\right\rangle^{\frac{2-s}{4}}
    \left\langle\|{\bm u}\|^2_{L^2}\right\rangle^\frac{s}{4}
    = \nu^{\frac{s-2}{4}}\ell^{-s}\bar{F} \chi^{\frac{2-s}{4}}U^\frac{s}{2}.
\end{align}
Rearranging yields a bound analogous to \eqref{eq:chi_inter} where now $F$ is replaced by $\bar{F}$\,:
\begin{equation}
\label{eq:chi-Fbar}
\chi \leqslant \nu^{-\frac{2-s}{2+s}} \ell^{-\frac{4s}{2+s}} \bar{F}^{\frac{4}{2+s}}  U^{\frac{2s}{2+s}}\,.
\end{equation}
To estimate $\bar{F}$ in terms of $U$, we multiply again the NSEs by a smooth divergence-free field $\bm\psi$, which must now depend on time. One option is $\bm\psi(\bm x,t)=F(t)\bm\Phi_G(\bm x,t)$, where $\bm \Phi_G$ is a suitable Galerkin truncation, in the space variables, of the shape function $\bm\Phi$ (the truncation set may now have to vary over time in such a way as to ensure that the space integral of $\bm\Phi\bcdot\bm\Phi_G$ remains positive for all $t$). Here, following \cite{df02}, we prefer to take
$\bm\psi=(-\widetilde{\Delta})^{-M}\bm f$, where $(-\widetilde{\Delta})^{-M}\bm f$ denotes the field whose Fourier coefficients are $(\ell k)^{-2M}\hat{\bm f_{\bm k}}$. This choice leads to a more natural interpretation of the pre-factors that appear in the estimates. The exponent $M$ must satisfy $M\geqslant 1-s/2$ for reasons that  will be clear immediately. After multiplying \eqref{eq:NS-u} by $\bm\psi$, integrating by parts, and time averaging, we obtain the analogous of \eqref{eq:tavf} and \eqref{eq:bound-F} (note that the term involving the time derivative of the velocity can no longer be ignored)
\begin{equation}
 \left\langle \int_{\Omega}\bm f \bcdot \bm \psi \, {\rm d}{\bm x} \right\rangle 
\leqslant \left|
\left\langle\int_\Omega {\bm u}\bcdot\partial_t{\bm\psi}\,{\rm d}{\bm x}\right\rangle\right|
+ \left|\left\langle\int_\Omega {\bm u}\bcdot\bnabla{\bm\psi}\bcdot{\bm u}\,{\rm d}{\bm x}\right\rangle\right|
    + \nu\,\left|\left\langle\int_\Omega {\bm u}\bcdot\Delta{\bm\psi}\,{\rm d}{\bm x}\right\rangle\right|
             \label{eq:FUt}
\end{equation}
with
\begin{align}
\left|
\left\langle\int_\Omega {\bm u}\bcdot\partial_t{\bm\psi}\,{\rm d}{\bm x}\right\rangle\right| 
&\leqslant \left\langle\Vert \partial_t\bm \psi \Vert_{L^2}^2\right\rangle^{1/2}  \left\langle \Vert \bm u\Vert _{L^2}^2 \right\rangle^{1/2}
         = \left\langle\Vert \partial_t (-\widetilde{\Delta})^{-M}\bm f \Vert_{L^2}^2\right\rangle^{1/2} U\ell\,,
         \label{eq:FUtbound1}
    \\ \left|\left\langle\int_\Omega {\bm u}\bcdot\bnabla{\bm\psi}\bcdot{\bm u}\,{\rm d}{\bm x}\right\rangle\right| 
         & \leqslant
         \big(\sup_t\Vert \bnabla \bm \psi\Vert _{L^{\infty}}\big) \left\langle \Vert \bm u\Vert _{L^2}^2 \right\rangle = \big(\sup_t\Vert \widetilde{\bnabla} (-\widetilde{\Delta})^{-M}\bm f\Vert _{L^{\infty}}\big) U^2\ell \,,
         \label{eq:FUtbound2}
         \\
	\left|\left\langle\int_\Omega {\bm u}\bcdot\Delta{\bm\psi}\,{\rm d}{\bm x}\right\rangle\right| 
	&\leqslant  \left\langle\Vert \Delta \bm \psi \Vert_{L^2}^2\right\rangle^{1/2}  \left\langle \Vert \bm u\Vert _{L^2}^2 \right\rangle^{1/2}
	 =  \left\langle\Vert \bm f \Vert_{\dot{H}^{2(1-M)}}^2\right\rangle^{1/2}  U\ell^{1-2M}\,.
    \label{eq:FUtbound3}
\end{align}
The constraint $M\geqslant 1-s/2$ ensures  the existence of $\Vert \bm f \Vert_{\dot{H}^{2(1-M)}}$ in the last term.
The left-hand side of \eqref{eq:FUt} can be bounded from below so as to make the dependence on $\bar{F}$ explicit\,:
\begin{equation}
 \left\langle \int_{\Omega}\bm f \bcdot \bm \psi \, {\rm d}{\bm x} \right\rangle = \ell^{-2M}
 \left\langle F^2 \Vert\bm \Phi\Vert^2_{\dot H^{-M}}
 \right\rangle \geqslant a^{-1} \ell^{2}\bar{F}^2 
 \label{eq:lower-bound}
\end{equation}
where
\begin{equation}
    a^{-1} = \ell^{-2(M+1)}\inf_t   \Vert\bm \Phi\Vert^2_{\dot H^{-M}}
\end{equation}
is positive, since $\ell^{s-1}\Vert\bm\Phi\Vert_{\dot H^s}=1$ for all $t$ and $-M\leqslant s$ for the range of values of $s$ considered here. Substituting the bounds in \eqref{eq:FUtbound1} to \eqref{eq:lower-bound} into \eqref{eq:FUt} and dividing through by $a^{-1}\ell^{2}\bar{F}$ gives
\begin{equation}
\bar{F}\leqslant 
a\omega_f U+
b_{1}\ell^{-1}U^{2} +  b_{2}\nu\ell^{-2} U\,,
\label{eq:bound-Fbar}
\end{equation}
where
\begin{equation}
\omega_f = \ell\bar{F}^{-1}\left\langle\Vert \partial_t (-\widetilde{\Delta})^{-M}\bm f\Vert_{L^2}^2\right\rangle^{1/2} 
\end{equation}
is a time frequency associated with the forcing and
\begin{equation}
    b_1 = a\bar{F}^{-1}\sup_t\Vert \widetilde{\bnabla} (-\widetilde{\Delta})^{-M}\bm f\Vert _{L^{\infty}}\,,
    \qquad
    b_2 = a\ell^{1-2M}\bar{F}^{-1}\left\langle\Vert \bm f \Vert_{\dot{H}^{2(1-M)}}^2\right\rangle^{1/2}\,.
\end{equation}
Note that $M$ can always be chosen large enough for $b_1$ to be bounded. We now use \eqref{eq:bound-Fbar} in \eqref{eq:chi-Fbar} to obtain
\begin{equation}
		\frac{\chi \ell^3}{U^3} 
		\leqslant \mathit{Re}^{\frac{2-s}{2+s}}\left( a \mathit{Sr}+b_1 + b_2 \mathit{Re}^{-1}\right)^{\frac{4}{2+s}},
\end{equation}
where 
\begin{equation}
\mathit{Sr}=\dfrac{\omega_f\ell}{U} 
\end{equation}
is the  Strouhal number, i.e.~the ratio of the convective time of the flow to the characteristic time scale of the forcing \citep[e.g.][]{horner_2002,voth_2003}. The pre-factor 
$b_2$ can be made independent of $F(t)$ by noting that
\begin{equation}
        b_2 \leqslant a\ell^{1-2M}\sup_t \Vert \bm \Phi \Vert_{\dot{H}^{2(1-M)}}\,.
\end{equation}
However, $b_1$ depends not only on the shape of the forcing, but also on the ratio of the maximum excursion of $F(t)$ to its root-mean-square value. It can be written as
\begin{equation}
    b_1 = a\sup_t\Vert \widetilde{\bnabla} \bm (-\widetilde{\Delta})^{-M}\bm\Phi\Vert _{L^{\infty}}\frac{\sup_t F(t)}{\bar{F}(t)}\,.
\end{equation}
Finally, by applying \eqref{eq:epsilon-chi}, we find the following bound for the time-averaged energy dissipation rate
\begin{equation}
	\frac{ \varepsilon \ell }{U^3}  \leqslant \mathit{Re}^{-\frac{s}{2+s}}\left(a\mathit{Sr+}b_1 + b_2 \mathit{Re}^{-1} \right)^{\frac{2}{2+s}}.
\end{equation}
Let us now consider the case $-1\leqslant s\leqslant 0$. The energy $\Vert \bm u\Vert_{L^2}$ is uniformly bounded in time even when the forcing depends on time \citep[to see this, simply replace $\Vert\bm f\Vert_{L^{2}}^2$ with $\sup_t\Vert\bm f\Vert_{L^{2}}^2$ in the derivation of][]{df02}. Therefore, $\varepsilon$  is estimated as in \eqref{eq:u-f}. We thus proceed as in \S\ref{sec:s<0} with an additional application of the Cauchy-Schwarz inequality to split the time average:
\begin{align}
    \varepsilon 
    & \nonumber
    \leqslant\ell^{-s-1}
    \left\langle F\,\|{\bm u}\|_{\dot H^{-s}}
    \right\rangle
    \leqslant \ell^{-s-1}\bar{F}
    \left\langle \|{\bm u}\|_{\dot H^{-s}}^2
    \right\rangle^{1/2}
    \leqslant \bar{F}\,\left\langle\|{\bm u}\|_{\dot H^1}^{-2s}\|{\bm u}\|^{2(1+s)}_{L^2}\right\rangle^{1/2}
    \\
    &\leqslant \ell^{-s-1}
    \bar{F}\,\left\langle\|{\bm u}\|_{\dot H^1}^{2}\right\rangle^{-\frac{s}{2}}
    \left\langle\|{\bm u}\|^2_{L^2}\right\rangle^\frac{1+s}{2}
    = \bar{F} \,\nu^\frac{s}{2} \ell^{-s}\varepsilon^{-\frac{s}{2}}U^{1+s}.
    \end{align}
This bound is the same as \eqref{eq:bound-cheskidov} with $F$ replaced by $\bar{F}$. Therefore, by using \eqref{eq:bound-Fbar} and rearranging, we find
\begin{equation}
\frac{\varepsilon \ell}{U^3} \leqslant \Rey^{-\frac{s}{2+s}}\big(a\mathit{Sr}+b_1 + b_2\,\Rey^{-1}\big)^{\frac{2}{2+s}}\,.
\end{equation}
In conclusion, for all $-1\leqslant s\leqslant 2$,  the bounds on the time-averaged dissipation rates retain the same form as in the case of constant-in-time forcing, the main difference being an additional $\mathit{Sr}$-dependent pre-factor that becomes relevant at large $\Rey$. Moreover, in general  $b_1$ now depends on features of the  forcing other than just the shape.

\section{Concluding remarks}\label{sect:conclusions}

The main result of our study is that the scenario predicted by \cite{ad06} and~\cite{gp07} for the two-dimensional NSEs only applies if the forcing is highly regular. By investigating the full range of forcing regularity (as allowed by the theory of weak solutions), we have shown that alternative scaling regimes are possible. Our results for constant-in-time forcing are summarized in table \ref{table}, in which only the large-$\Rey$ bounds are shown. Leaving aside the special case of single-$k$ or constant-$\varepsilon$ forcing, three distinct regimes can be identified as the degree of regularity of the forcing is varied. The existence of these three regimes only becomes apparent when the estimates in terms of $\Gr$ are converted into an $\Rey$-dependent form.
\par\smallskip
For $s\geqslant 2$, the estimates obtained by \cite{ad06} and \cite{gp07} are consistent with the dual-cascade scenario. In the second regime ($0 <  s < 2$), a direct cascade of energy must still be ruled out in the infinite-$\Rey$ limit. However, the amount of energy injected at small scales increases as $s$ decreases, so that $\varepsilon$ decreases monotonically as a function of $\Rey$, but increasingly slowly. In this regime, $\chi$ is allowed to grow with $\Rey$. As energy injection at small scales increases, the bound on the number of degrees of freedom of the system (quantified by the attractor dimension) also increases from $\Rey$ for $s=2$ to $\Rey^{4/3}$ for $s=0$. Correspondingly, the scaling of the smallest length scale in the flow varies between $\eta_\chi\sim\ell\Rey^{-1/2}$ and $\eta_\chi \sim \ell\Rey^{-2/3}$, for the same range of $s$. Finally, for $-1 < s < 0$, the time-averaged energy dissipation rate can increase with $\Rey$, similar to what is found in three dimensions for non-square-integrable forcing \citep{mazzi2004fractal,cdp07}. For all $-1\leqslant s\leqslant 2$, if the forcing is time-dependent, the large-$\Rey$ estimates of the mean dissipation rates are multiplied by a $\mathit{Sr}$-dependent pre-factor.
\par\smallskip
It is noteworthy that three dynamical regimes are also observed in the two-dimensional NSEs with power-law {\it stochastic} forcing \citep{mazzino2007PRL,mazzino2009IOP}. The covariance of the body force in Fourier space is taken proportional to $k^{2-y}$ with $y\geqslant 0$. If $y>6$, the enstrophy flux is constant, and a direct cascade of enstrophy is observed, which is consistent with the KLB theory. In the range $4< y  < 6$, there is scale-by-scale balance of the energy and enstrophy fluxes. In addition, the enstrophy flux increases with $k$, whereas the energy flux decays. For $0\leqslant y < 4$, the system displays an inverse energy cascade. While some similarities with our results may exist for the two regimes with more regular forcing, a systematic comparison of the two systems may not be appropriate. Indeed, our forcing is deterministic, whereas in \cite{mazzino2007PRL,mazzino2009IOP} the forcing is stochastic and the analysis explicitly uses its Gaussian and white-in-time nature. Moreover, the range $0\leqslant y<2$ corresponds to highly irregular body forces that are not covered by our study.
\par\smallskip
It is also interesting to note that the lower bound on the mean square velocity  in \eqref{eq:F_amp} holds independently of the degree of regularity of $\bm f$. Therefore, if the forcing is constant in time and the Grashof number is defined in terms of the $\dot H^s$-norm of $\bm f$ as $\Gr = \ell^3 \nu^{-2}F$, then the Doering and Foias relation between $\Gr$ and $\Rey$ does not depend on $s$, and satisfies (\ref{DF02}) for all $s\geqslant-1$. If the forcing is time-dependent, it is appropriate to define  $\Gr$ in terms of the root-mean-square of the magnitude of the forcing as $\Gr = \ell^3\nu^{-2}\bar{F}$. In this case, \eqref{eq:bound-Fbar} gives
\begin{equation}
    \Gr \leqslant (a\mathit{Sr}+b_1) \Rey^2 + b_2 \Rey\,.
\end{equation}
Another implication of \eqref{eq:F_amp} and \eqref{eq:bound-Fbar} is the existence of lower bounds for $\varepsilon$ and $\chi$ at large $\Gr$ (similar to a lower bound for $\varepsilon$ found by \citealt{df02} in the case of square-integrable forcing). Indeed, Poincar\'e's inequality gives 
\begin{equation}
    \varepsilon \geqslant 4\pi^2\nu \ell^{-2} U^2, \qquad
    \chi \geqslant 16\pi^4\nu\ell^{-4} U^2.
\end{equation}
These bounds imply that the rescaled dissipation rates are both bounded below by $\Rey^{-1}$ \citep{ad06}. While this result captures the laminar behaviour, it is of little use in the turbulent regime. To obtain results that are relevant at large $\Gr$ (and hence at large $\Rey$), we observe that \eqref{eq:F_amp} and \eqref{eq:bound-Fbar} can be rewritten as
\begin{equation}
\kappa_1 \nu^{-2}\ell^2\Gr^{-1} U^2 +\kappa_2 \nu^{-1}\ell\Gr^{-3/2}U-1\geqslant 0\,,
\end{equation}
where $\kappa_1=c_1$ and $\kappa_2=c_2$ for constant-in-time forcing, while $\kappa_1=a\mathit{Sr}+b_1$ and $\kappa_2=b_2$ for time-dependent forcing. Therefore, as $\Gr\to\infty$,
\begin{equation}
    U^2 \geqslant \kappa_1^{-1} \nu^2\ell^{-2}\Gr
\end{equation}
and
\begin{equation}
    \varepsilon \geqslant \dfrac{4\pi^2}{\kappa_1}\, \nu^3\ell^{-4} \Gr\,, \qquad
    \chi \geqslant \dfrac{16\pi^4}{\kappa_1}\,\nu^{3}\ell^{-6} \Gr\,.
\end{equation}
The form of these bounds is independent of the degree of regularity of the forcing, which only enters the definition of $\Gr$. If the forcing is time-dependent, the pre-factor depends on the Strouhal number and, at large $\mathit{Sr}$, both lower bounds are $O\left(\mathit{Sr}^{-1}\Gr\right)$.

\begin{bmhead}[Acknowledgements.]
Thanks are due to Franco Flandoli (Scuola Normale Superiore), Giorgio Krstulovic (INPHYNI) and Samriddhi Sankar Ray (ICTS--TIFR) for insightful discussions.  R.~M. is grateful for the hospitality of Laboratoire Jean Alexandre Dieudonn\'e, Universit\'e C\^ote d'Azur, where part of this work was performed. 
\end{bmhead}

\begin{bmhead}[Funding.]
This work was supported by the French government through the UCA$^{\rm JEDI}$ Investments in the Future project managed by the National Research Agency
(ANR) with reference number ANR-15-IDEX-01.
\end{bmhead}


\bibliographystyle{jfm_arxiv}
\bibliography{refs}

\begin{thebibliography}{56}
\expandafter\ifx\csname natexlab\endcsname\relax\def\natexlab#1{#1}\fi
\def\au#1{#1} \def\ed#1{#1} \def\yr#1{#1}\def\at#1{#1}\def\jt#1{\textit{#1}}
  \def\bt#1{#1}\def\bvol#1{\textbf{#1}} \def\vol#1{#1} \def\pg#1{#1}
  \def\publ#1{#1}\def\arxiv#1{#1}\def\org#1{#1}\def\st#1{\textit{#1}}

\bibitem[Adzhemyan {\em et~al.\/}(1999)Adzhemyan, Antonov \&
  Vasiliev]{adzhemyan1999field}
{\normalfont \au{Adzhemyan, L.~Ts.}, \au{Antonov, N.~V.} \& \au{Vasiliev,
  A.~N.}} \yr{1999} {\em Field Theoretic Renormalization Group in Fully
  Developed Turbulence\/}.  \publ{CRC Press}.

\bibitem[Alexakis \& Biferale(2018)]{ab18}
{\normalfont \au{Alexakis, A.} \& \au{Biferale, L.}} \yr{2018}  \at{Cascades
  and transitions in turbulent flows}.  \jt{Phys. Rep.}  \bvol{767--769},
  \pg{1--102}.

\bibitem[Alexakis \& Doering(2006)]{ad06}
{\normalfont \au{Alexakis, A.} \& \au{Doering, C.~R.}} \yr{2006}  \at{Energy
  and enstrophy dissipation in steady state 2d turbulence}.  \jt{Phys. Lett.
  {\rm A}}  \bvol{359},  \pg{652--657}.

\bibitem[Bardos \& Titi(2013)]{bt13}
{\normalfont \au{Bardos, C.~W.} \& \au{Titi, E.~S.}} \yr{2013}  \at{Mathematics
  and turbulence: where do we stand?}  \jt{J. Turbul.}  \bvol{14},
  \pg{42--76}.

\bibitem[Bedrossian \& Vicol(2022)]{vicol2022}
{\normalfont \au{Bedrossian, J.} \& \au{Vicol, V.}} \yr{2022} {\em The
  Mathematical Analysis of the Incompressible Euler and {Navier--Stokes}
  Equations\/}.  \publ{Providence, RI: American Mathematical Society}.

\bibitem[Benzi \& Toschi(2023)]{bt23}
{\normalfont \au{Benzi, R.} \& \au{Toschi, F.}} \yr{2023}  \at{Lectures on
  turbulence}.  \jt{Phys. Rep.}  \bvol{1021},  \pg{1--106}.

\bibitem[Biferale {\em et~al.\/}(2004{\natexlab{{\em a\/}}})Biferale, Cencini,
  Lanotte, Sbragaglia \& Toschi]{Biferale_2004}
{\normalfont \au{Biferale, L.}, \au{Cencini, M.}, \au{Lanotte, A.~S.},
  \au{Sbragaglia, M.} \& \au{Toschi, F.}} \yr{2004{\natexlab{{\em a\/}}}}
  \at{Anomalous scaling and universality in hydrodynamic systems with power-law
  forcing}.  \jt{New J. Phys.}  \bvol{6},  \pg{37}.

\bibitem[Biferale {\em et~al.\/}(2004{\natexlab{{\em b\/}}})Biferale, Lanotte
  \& Toschi]{biferale2004}
{\normalfont \au{Biferale, L.}, \au{Lanotte, A.~S.} \& \au{Toschi, F.}}
  \yr{2004{\natexlab{{\em b\/}}}}  \at{Effects of forcing in three-dimensional
  turbulent flows}.  \jt{Phys. Rev. Lett.}  \bvol{92},  \pg{094503}.

\bibitem[Boffetta \& Ecke(2012)]{be12}
{\normalfont \au{Boffetta, G.} \& \au{Ecke, R.~E.}} \yr{2012}
  \at{Two-dimensional turbulence}.  \jt{Annu. Rev. Fluid Mech.}  \bvol{44},
  \pg{427--451}.

\bibitem[Canet(2022)]{canet_2022}
{\normalfont \au{Canet, L.}} \yr{2022}  \at{Functional renormalisation group
  for turbulence}.  \jt{J. Fluid Mech.}  \bvol{950},  \pg{P1}.

\bibitem[Cheskidov {\em et~al.\/}(2007)Cheskidov, Petrov \& Doering]{cdp07}
{\normalfont \au{Cheskidov, A.}, \au{Petrov, N.} \& \au{Doering, C.~R.}}
  \yr{2007}  \at{Energy dissipation in fractal-forced flow}.  \jt{J. Math.
  Phys.}  \bvol{48},  \pg{065208}.

\bibitem[Constantin(2006)]{constantin2006}
{\normalfont \au{Constantin, P.}} \yr{2006}  \at{{Euler Equations,
  Navier--Stokes Equations and Turbulence}}.  \bt{In {\em Mathematical
  Foundation of Turbulent Viscous Flows\/} (ed. \ed{M.~Cannone \&
  T.~Miyakawa})},  \st{Lecture Notes in Mathematics},  \vol{vol. 1871},
  \pg{pp. 1--43}.  \publ{Berlin Heidelberg: Springer}.

\bibitem[Constantin {\em et~al.\/}(1994)Constantin, Foias \& Manley]{cfm94}
{\normalfont \au{Constantin, P.}, \au{Foias, C.} \& \au{Manley, O.~P.}}
  \yr{1994}  \at{Effects of the forcing function spectrum on the energy
  spectrum in 2-{\rm d} turbulence}.  \jt{Phys. Fluids}  \bvol{6},
  \pg{427--429}.

\bibitem[Constantin {\em et~al.\/}(1988)Constantin, Foias \& Temam]{cft88}
{\normalfont \au{Constantin, P.}, \au{Foias, C.} \& \au{Temam, R.}} \yr{1988}
  \at{On the dimension of the attractors in two-dimensional turbulence}.
  \jt{Physica {\rm D}}  \bvol{30},  \pg{284--296}.

\bibitem[Dascaliuc {\em et~al.\/}(2008)Dascaliuc, Foias \& Jolly]{dfj08}
{\normalfont \au{Dascaliuc, R.}, \au{Foias, C.} \& \au{Jolly, M.S.}} \yr{2008}
  \at{Some specific mathematical constraints on {2D} turbulence}.  \jt{Physica
  {\rm D}}  \bvol{237},  \pg{3020--3029}.

\bibitem[De {\em et~al.\/}(2023)De, Mitra \& Pandit]{de2023dynamic}
{\normalfont \au{De, S.}, \au{Mitra, D.} \& \au{Pandit, R.}} \yr{2023}
  \at{Dynamic multiscaling in stochastically forced {B}urgers turbulence}.
  \jt{Sci. Rep.}  \bvol{13},  \pg{7151}.

\bibitem[Doering(1999)]{doering1999}
{\normalfont \au{Doering, C.~R.}} \yr{1999}  \at{Does turbulence saturate
  global transport estimates?}  \bt{In {\em Fundamental Problematic Issues in
  Turbulence\/} (ed. \ed{A.~Gyr, W.~Kinzelbach \& A.~Tsinober})},  \pg{pp.
  3--16}.  \publ{Basel: Springer}.

\bibitem[Doering(2009)]{doering2009}
{\normalfont \au{Doering, C.~R.}} \yr{2009}  \at{The 3{\rm d}
  {N}avier--{S}tokes problem}.  \jt{Annu. Rev. Fluid Mech.}  \bvol{41},
  \pg{109--128}.

\bibitem[Doering \& Foias(2002)]{df02}
{\normalfont \au{Doering, C.~R.} \& \au{Foias, C.}} \yr{2002}  \at{Energy
  dissipation in body-forced turbulence}.  \jt{J. Fluid Mech.}  \bvol{467},
  \pg{289--306}.

\bibitem[Doering \& Gibbon(1991)]{dg91}
{\normalfont \au{Doering, C.~R.} \& \au{Gibbon, J.~D.}} \yr{1991}  \at{Note on
  the {Constantin--Foias--Temam} attractor dimension estimate for
  two-dimensional turbulence}.  \jt{Physica {\rm D}}  \bvol{48},
  \pg{471--480}.

\bibitem[Doering \& Gibbon(1995)]{dg95}
{\normalfont \au{Doering, C.~R.} \& \au{Gibbon, J.~D.}} \yr{1995} {\em Applied
  Analysis of the {N}avier--{S}tokes Equations\/}.  \publ{Cambridge: Cambridge
  University Press}.

\bibitem[Eyink(1996)]{eyink1996exact}
{\normalfont \au{Eyink, G.~L.}} \yr{1996}  \at{Exact results on stationary
  turbulence in 2d: consequences of vorticity conservation}.  \jt{Physica {\rm
  D}}  \bvol{91},  \pg{97--142}.

\bibitem[Foias(1997)]{foias_1997}
{\normalfont \au{Foias, C.}} \yr{1997}  \at{What do the {Navier--Stokes}
  equations tell us about turbulence?}  \jt{Contemp. Math.}  \bvol{208},
  \pg{151--180}.

\bibitem[Foias {\em et~al.\/}(2002)Foias, Jolly, Manley \& Rosa]{fjmr02}
{\normalfont \au{Foias, C.}, \au{Jolly, M.~S.}, \au{Manley, O.~P.} \& \au{Rosa,
  R.}} \yr{2002}  \at{Statistical estimates for the {Navier--Stokes} equations
  and the {Kraichnan} theory of 2-d fully developed turbulence}.  \jt{J. Stat.
  Phys.}  \bvol{108},  \pg{591--645}.

\bibitem[Foias {\em et~al.\/}(2001)Foias, Manley, Rosa \& Temam]{fmrt01}
{\normalfont \au{Foias, C.}, \au{Manley, O.}, \au{Rosa, R.} \& \au{Temam, R.}}
  \yr{2001} {\em Navier--Stokes Equations and Turbulence\/}.  \publ{Cambridge:
  Cambridge University Press}.

\bibitem[Foias {\em et~al.\/}(1993)Foias, Manley \&
  Temam]{foiasmanleytemam1993}
{\normalfont \au{Foias, C.}, \au{Manley, O.~P.} \& \au{Temam, R.}} \yr{1993}
  \at{Bounds for the mean dissipation of 2-{\rm d} enstrophy and 3-{\rm d}
  energy in turbulent flows}.  \jt{Phys. Lett. {\rm A}}  \bvol{174},
  \pg{210--215}.

\bibitem[Foias {\em et~al.\/}(1983)Foias, Manley, Temam \& Treve]{fmtt83}
{\normalfont \au{Foias, C.}, \au{Manley, O.~P.}, \au{Temam, R.} \& \au{Treve,
  Y.~M.}} \yr{1983}  \at{Asymptotic analysis of the {Navier--Stokes}
  equations}.  \jt{Physica {\rm D}}  \bvol{9},  \pg{157--188}.

\bibitem[Frisch(1995)]{frisch1995}
{\normalfont \au{Frisch, U.}} \yr{1995} {\em Turbulence: The Legacy of
  A.~N.~Kolmogorov\/}.  \publ{Cambridge: Cambridge University Press}.

\bibitem[Gibbon(2019)]{gibbon19}
{\normalfont \au{Gibbon, J.~D.}} \yr{2019}  \at{Weak and strong solutions of
  the {3$D$} {Navier--Stokes} equations and their relation to a chessboard of
  convergent inverse length scales}.  \jt{J. Nonlin. Sci.}  \bvol{29},
  \pg{215--228}.

\bibitem[Gibbon(2020)]{gibbon2020epl}
{\normalfont \au{Gibbon, J.~D.}} \yr{2020}  \at{Intermittency, cascades and
  thin sets in three-dimensional {Navier--Stokes} turbulence}.  \jt{Europhys.
  Lett.}  \bvol{131},  \pg{64001}.

\bibitem[Gibbon \& Pavliotis(2007)]{gp07}
{\normalfont \au{Gibbon, J.~D.} \& \au{Pavliotis, G.~A.}} \yr{2007}
  \at{Estimates for the two-dimensional {N}avier--{S}tokes equations in terms
  of the {R}eynolds number}.  \jt{J. Math. Phys.}  \bvol{48},  \pg{065202}.

\bibitem[Horner {\em et~al.\/}(2002)Horner, Metcalfe, Wiggins \&
  Ottino]{horner_2002}
{\normalfont \au{Horner, M.}, \au{Metcalfe, G.}, \au{Wiggins, S.} \&
  \au{Ottino, J.~M.}} \yr{2002}  \at{Transport enhancement mechanisms in open
  cavities}.  \jt{J. Fluid Mech.}  \bvol{452},  \pg{199--229}.

\bibitem[Ladyzhenskaya(1975)]{lad75}
{\normalfont \au{Ladyzhenskaya, O.~A.}} \yr{1975}  \at{A dynamical system
  generated by the {Navier--Stokes} equations}.  \jt{J. Sov. Math.}  \bvol{3},
  \pg{458–479}.

\bibitem[Lopes~Filho {\em et~al.\/}(2006)Lopes~Filho, Mazzucato \&
  Nussenzveig~Lopes]{mazzuccato06}
{\normalfont \au{Lopes~Filho, M.}, \au{Mazzucato, A.} \& \au{Nussenzveig~Lopes,
  H.}} \yr{2006}  \at{Renormalized solutions and enstrophy defects in {2D}
  turbulence}.  \jt{Arch. Ration. Mech. Anal.}  \bvol{179},  \pg{353–387}.

\bibitem[Matharu {\em et~al.\/}(2022)Matharu, Protas \& Yoneda]{mpy22}
{\normalfont \au{Matharu, P.}, \au{Protas, B.} \& \au{Yoneda, T.}} \yr{2022}
  \at{On maximum enstrophy dissipation in {2D} {Navier–Stokes} flows in the
  limit of vanishing viscosity}.  \jt{Physica {\rm D}}  \bvol{441},
  \pg{133517}.

\bibitem[Mazzi {\em et~al.\/}(2002)Mazzi, Okkels \& Vassilicos]{mazzi2002}
{\normalfont \au{Mazzi, B.}, \au{Okkels, F.} \& \au{Vassilicos, J.~C.}}
  \yr{2002}  \at{A shell-model approach to fractal-induced turbulence}.
  \jt{Eur. Phys. J. {\rm B}}  \bvol{28},  \pg{243--251}.

\bibitem[Mazzi \& Vassilicos(2004)]{mazzi2004fractal}
{\normalfont \au{Mazzi, B} \& \au{Vassilicos, J.~C.}} \yr{2004}
  \at{Fractal-generated turbulence}.  \jt{J. Fluid Mech.}  \bvol{502},
  \pg{65--87}.

\bibitem[Mazzino {\em et~al.\/}(2007)Mazzino, Muratore-Ginanneschi \&
  Musacchio]{mazzino2007PRL}
{\normalfont \au{Mazzino, A.}, \au{Muratore-Ginanneschi, P.} \& \au{Musacchio,
  S.}} \yr{2007}  \at{Scaling properties of the two-dimensional randomly
  stirred {N}avier--{S}tokes equation}.  \jt{Phys. Rev. Lett.}  \bvol{99},
  \pg{144502}.

\bibitem[Mazzino {\em et~al.\/}(2009)Mazzino, Muratore-Ginanneschi \&
  Musacchio]{mazzino2009IOP}
{\normalfont \au{Mazzino, A.}, \au{Muratore-Ginanneschi, P.} \& \au{Musacchio,
  S.}} \yr{2009}  \at{Scaling regimes of 2d turbulence with power-law stirring:
  theories versus numerical experiments}.  \jt{J. Stat. Mech.}  \bvol{2009},
  \pg{P10012}.

\bibitem[Mitra {\em et~al.\/}(2005)Mitra, Bec, Pandit \&
  Frisch]{mitra2005stochastic}
{\normalfont \au{Mitra, D.}, \au{Bec, J.}, \au{Pandit, R.} \& \au{Frisch, U.}}
  \yr{2005}  \at{Is multiscaling an artifact in the stochastically forced
  {B}urgers equation?}  \jt{Phys. Rev. Lett.}  \bvol{94},  \pg{194501}.

\bibitem[Monin \& Yaglom(1975)]{my75}
{\normalfont \au{Monin, A.~S.} \& \au{Yaglom, A.~M.}} \yr{1975} {\em
  Statistical Fluid Mechanics\/}.  \publ{Cambridge, MA: MIT Press}.

\bibitem[Robinson(1995)]{robinson1995finite}
{\normalfont \au{Robinson, J.~C.}} \yr{1995}  \at{Finite-dimensional behavior
  in dissipative partial differential equations}.  \jt{Chaos}  \bvol{5},
  \pg{330--345}.

\bibitem[Robinson(2001)]{robinson2001}
{\normalfont \au{Robinson, J.~C.}} \yr{2001} {\em Infinite-Dimensional
  Dynamical Systems\/}.  \publ{Cambridge: Cambridge University Press}.

\bibitem[Robinson(2003)]{robinson2003}
{\normalfont \au{Robinson, J.~C.}} \yr{2003}  \at{Low dimensional attractors
  arise from forcing at small scale}.  \jt{Physica {\rm D}}  \bvol{181},
  \pg{39--44}.

\bibitem[Robinson(2007)]{robinson2007}
{\normalfont \au{Robinson, J.~C.}} \yr{2007}  \at{Parametrization of global
  attractors, experimental observations, and turbulence}.  \jt{J. Fluid Mech.}
  \bvol{578},  \pg{495–507}.

\bibitem[Robinson(2013)]{robinson2013}
{\normalfont \au{Robinson, J.~C.}} \yr{2013}  \at{Attractors and
  finite-dimensional behaviour in the {2D} {Navier--Stokes} equations}.
  \jt{Int. Sch. Res. Notices}  \bvol{2013},  \pg{291823}.

\bibitem[Robinson {\em et~al.\/}(2016)Robinson, Rodrigo \&
  Sadowski]{robinson2016}
{\normalfont \au{Robinson, J.~C.}, \au{Rodrigo, J.~L.} \& \au{Sadowski, W.}}
  \yr{2016} {\em The Three-Dimensional Navier--Stokes Equations\/}.
  \publ{Cambridge: Cambridge University Press}.

\bibitem[Sain {\em et~al.\/}(1998)Sain, Manu \& Pandit]{sain1998}
{\normalfont \au{Sain, A.}, \au{Manu} \& \au{Pandit, R.}} \yr{1998}
  \at{Turbulence and multiscaling in the randomly forced {N}avier--{S}tokes
  equation}.  \jt{Phys. Rev. Lett.}  \bvol{81},  \pg{4377}.

\bibitem[Smith \& Woodruff(1998)]{smith1998renormalization}
{\normalfont \au{Smith, L.~M.} \& \au{Woodruff, S.~L.}} \yr{1998}
  \at{Renormalization-group analysis of turbulence}.  \jt{Annu. Rev. Fluid
  Mech.}  \bvol{30},  \pg{275--310}.

\bibitem[Tran \& Dritschel(2006)]{td06}
{\normalfont \au{Tran, C.~V.} \& \au{Dritschel, D.~G.}} \yr{2006}
  \at{Vanishing enstrophy dissipation in two-dimensional {N}avier--{S}tokes
  turbulence in the inviscid limit}.  \jt{J. Fluid Mech.}  \bvol{559},
  \pg{107–116}.

\bibitem[Tran \& Shepherd(2002)]{tran2002constraints}
{\normalfont \au{Tran, C.~V.} \& \au{Shepherd, T.~G.}} \yr{2002}
  \at{Constraints on the spectral distribution of energy and enstrophy
  dissipation in forced two-dimensional turbulence}.  \jt{Physica {\rm D}}
  \bvol{165},  \pg{199--212}.

\bibitem[Tran {\em et~al.\/}(2004)Tran, Shepherd \& Cho]{tsc04}
{\normalfont \au{Tran, C.~V.}, \au{Shepherd, T.~G.} \& \au{Cho, H.-R.}}
  \yr{2004}  \at{Extensivity of two-dimensional turbulence}.  \jt{Physica {\rm
  D}}  \bvol{192},  \pg{187--195}.

\bibitem[Vassilicos(2015)]{vassilicos2015}
{\normalfont \au{Vassilicos, J~Christos}} \yr{2015}  \at{Dissipation in
  turbulent flows}.  \jt{Annu. Rev. Fluid Mech.}  \bvol{47},  \pg{95--114}.

\bibitem[Verma(2019)]{verma2019}
{\normalfont \au{Verma, M.}} \yr{2019} {\em Energy Transfers in Fluid Flows\/}.
   \publ{Cambridge: Cambridge University Press}.

\bibitem[Voth {\em et~al.\/}(2003)Voth, Saint, Dobler \& Gollub]{voth_2003}
{\normalfont \au{Voth, G.~A.}, \au{Saint, T.~C.}, \au{Dobler, G.} \&
  \au{Gollub, J.~P.}} \yr{2003}  \at{Mixing rates and symmetry breaking in
  two-dimensional chaotic flow}.  \jt{Phys. Fluids}  \bvol{15},
  \pg{2560--2566}.

\bibitem[Zhou(2010)]{zhou2010renormalization}
{\normalfont \au{Zhou, Y.}} \yr{2010}  \at{Renormalization group theory for
  fluid and plasma turbulence}.  \jt{Phys. Rep.}  \bvol{488},  \pg{1--49}.

\end{thebibliography}

\end{document}